\documentclass[12pt]{iopart}

\usepackage{color}
\usepackage{ulem}
\usepackage{graphicx}
\expandafter\let\csname equation*\endcsname\relax      
\expandafter\let\csname endequation*\endcsname\relax   
\usepackage{amsmath} 
\usepackage{amssymb} 
\usepackage{dcolumn}
\usepackage{bm}
\usepackage[utf8]{inputenc}
\usepackage[T1]{fontenc}
\usepackage{mathptmx}
\usepackage{multirow}
\usepackage{tikz}
\usepackage{hyperref}  
\begin{document}

\title[]{Power dependence of density limit due to plasma-wall interaction in a burning plasma}

\author{Jiaxing Liu$^{1}$, Ping Zhu$^{1,2*}$, Dominique Franck Escande$^{3}$}
\address{
{\small $^1$State Key Laboratory of Advanced Electromagnetic Technology,
International Joint Research Laboratory of Magnetic Confinement Fusion
and Plasma Physics, School of Electrical and Electronic Engineering,
Huazhong University of Science and Technology, Wuhan, 430074, China.} \\
{\small $^2$Department of Nuclear Engineering and Engineering Physics, University of Wisconsin-Madison, Madison, Wisconsin, 53706, United States of America.}\\
{\small $^3$Aix-Marseille Universit$\acute{\text{e}}$, CNRS, PIIM, UMR 7345, Marseille, France.}
}
\newcommand{\eadss}[1]{\address{E-mail: #1}}
\eadss{zhup@hust.edu.cn}
\vspace{10pt}
\begin{indented}
    \item[]\today
\end{indented}

\begin{abstract}
    The density limit is one of the major obstacles to achieving the desired fusion performance  in tokamaks. However, the underlying physics mechanism for its recently observed power dependence in experiments has not been well understood or predicted in theory.
    In this work, the power dependent scalings of density limit are obtained based on the plasma-wall self-organization theory [D.F. Escande 2022 NF], which are able to match the power dependence of density limits in multiple tokamak devices. The key factors influencing the power dependence are found to be the plasma-wall sputtering and the particle confinement time. The effects of non-sputtered impurities and fusion products are further evaluated. This PWSO-density limit model is then extended to the burning plasma regime and used to predict the conditions for entering burning plasma.
\end{abstract}

%
\vspace{2pc}
\noindent{\it Keywords}: Density limit, impurity radiation, plasma-wall interaction, burning plasma condition

%
%
%

\section{Introduction}
The operation of tokamak often terminates due to disruptions above an upper limit on plasma density $n_e$ \cite{Greenwald_1988,Greenwald_2002}, which significantly constrains the tokamak performance and ultimately influences its ability to achieve fusion ignition.
Experimentally, the well known Greenwald scaling law for density limit (DL) $\displaystyle
    n_{\text{\tiny G}}\left(10^{20}\text{m}^{-3}\right)={I_\text{p}\left(\text{MA}\right)}/{\pi a\left(\text{m}\right)^2}$ has been established for decades, where $I_\text{p}$ is the plasma current and $a$ is the minor radius \cite{Greenwald_1988,Greenwald_2002}.
However, many experiments demonstrate that the density limit increases with the heating power \cite{Zanca_2019} in various devices, such as JT-60\cite{Kamada_1991}, ASDEX-U\cite{Mertens_1997,Manz2023} and TEXTOR-96\cite{Rapp_1999}, where the exact power scalings differ in different heating regimes. For instance, DL is proportional to (independent of) heating power at lower (higher) heating power regime as shown in the Fig.~10 of Ref.~\cite{huber2017}.

Many experiments suggest that the density limit may be attributed to the cooling of tokamak boundary due to impurity radiation \cite{Greenwald_1988,Greenwald_2002,Mertens1994,Gates_2012,Hong2017}.
On the other hand, the Greenwald DL scaling is power independent, which is inconsistent with the above scenario of tokamak boundary cooling, because more heating power should be able to suppress or slow down the cooling process and the disruption.
Several theoretical models have since been developed to account for the heating power dependence of DL~\cite{Zanca_2019,Bernert2015,Eich2021,Brown2021,Giacomin2022,Singh2022,Stroth2022,Diamond2023,Manz2023}. For example, a power-balance model including the radiation losses from impurities and neutrals is able to predict the power-dependent DL scaling of many L-mode tokamak plasmas $n_c\propto P^{4/9}I_p^{4/9}$ \cite{Zanca_2019}, wich agrees with a large published data base, where $P$ is the input power to the system. 

The recent development of the plasma-wall self-organization (PWSO) theory may provide an alternative and new perspective on the power-dependence of the density limit in tokamak plasmas \cite{escande2022,jxliu2023}. In this work, we derive the heating power scaling of density limit due to the plasma wall interaction in a magnetically confined plasma based on the PWSO theory, where the target region plasma temperature $T_t$ depends on the power $P_t$ deposited on the wall, and ultimately the heating power.
For the burning plasma regime, the PWSO theory and the corresponding power-dependent DL are applied to re-evaluate the ignition condition as well as the impact from helium ash.

The remainder of this paper is organized as follows: section 2 derives the general power dependent density limit scaling based on 0D PWSO model. Section 3 gives the more specific power-dependent DL scaling based on the particular tokamak regimes, which are compared with experimental results and other theoretical DL scalings. Section 4 evaluate the effect of radiation due to non-sputtered impurities, such as the helium ash from fusion reactions. Section 5 offers a discussion about the effect of plasma-wall interaction on the ignition condition. Lastly, section 6 concludes with a summary and discussion.

\section{Effects of heating power on density limit in 0D PWSO model}\label{sec2}
The effect of heating power on density limit can be derived from the PWSO theory.
The plasma-wall self-organization leads to an upper density limit of a tokamak which is related to the sputtering properties of the wall's and target's material \cite{escande2022,jxliu2023}.
The radiation power due to sputtered impurity at next cycle time is determined by the properties of the wall material and the power deposited on the wall based on the assumption radiation distributes uniformly in the plasma, as shown in following
\begin{align}
    \label{eq:basicPWSO}
R_{+}=\int_VR_\text{imp,coe}n_en_{\text{imp,sputtered}}\mathrm{d}V&=R_\text{imp,coe}n_e\frac{f_\mathrm{ion} \lambda aI\left(T_t\right)}{{2 D_\perp T_t}} P_t=\alpha_1\frac{I(T_t)}{T_t} P_t \\
    \label{eq:pbe0}
    P_t&=P_\text{heat}-R    
\end{align}
where $P_t$ and $P_\text{heat}$ are the power deposition on wall and the total input power respectively at current cycle time, $R$ and $R_+$ are total radiation powers due to sputtered impurities at the current and next cycle time respectively, $R_\text{imp,coe}$ is the impurity radiation coefficient, and $\alpha_1={f_\mathrm{ion}\lambda n_eR_\text{imp,coe}a}/{(2D_\perp)}$. In the derivation of Eq.~(\ref{eq:basicPWSO}), the 0D model for the sputtered impurity density is used \cite{escande2022, stangeby2000}
\begin{equation}
    n_\text{imp,sputtered}={f_\mathrm{ion} \lambda P_t I\left(T_t\right)}/\left({2 \pi a L D_\perp T_t}\right)
\end{equation}
where $f_\mathrm{ion}$ represents the ionization rate of sputtered neutral impurity, $\lambda$ the distance between the target/wall and ionization location, $a$ the minor radius, $L$ the toroidal circumference, and $D_\perp$ the impurities' particle transport coefficient, $I\left(T_\text{t}\right)$
is the average of the yield function of carbon $Y\left(E\right)$ over the impinging particle energies.
\begin{equation}
  \label{eq:ITt}
  I\left(T_\text{t}\right)=\sqrt{\frac{m}{2\pi T_\text{t}}} \int_{0}^{\infty} Y\left(\frac{m v^{2}}{2}+\gamma T_\text{t}\right) \exp \frac{-m v^{2}}{2T_\text{t}} \mathrm{~d} v
\end{equation}
where  $\gamma$ is the total energy transmission coefficient \cite{stangeby2000}, $\gamma T_t$ is a measure of the Debye shield length, and $m$ is the ion mass of impinging particle.

Based on the relation between the target region plasma temperature $T_t$ and the power $P_t$~\cite{stangeby2000}, which we discuss about in detail in section~\ref{sec3}, the relation between power $P_t$ and sputtering property $I\left(T_t\right)/T_t$ can be established as follows
\begin{equation}
    \label{eq:FPtTt}
    \frac{I\left(T_t\right)}{T_t}=\frac{I\left(T_t(P_t)\right)}{T_t(P_t)}\equiv F\left(P_t\right)
\end{equation}
The steady state of the global power balance in Eq.~(\ref{eq:basicPWSO}) can be thus written as 
\begin{equation}
    \label{eq:radeq}
    \begin{aligned}
        R_*=\alpha_1\left(P_\text{heat}-R_*\right)F
    \end{aligned}
\end{equation}
The fraction of sputtered impurities is also assumed to have reached a steady state, which is represented by 
\begin{equation}
\label{eq:fimpPheat}
    f_\text{imp}=\frac{R_*}{n_c^2R_\text{coe,imp}}
\end{equation}
From the stable condition for the fixed point of Eq.~(\ref{eq:basicPWSO})
\begin{equation}
    \left|\frac{\partial R_+}{\partial R}\right|_{R=R_*}=\alpha\le1
\end{equation}
where $\alpha=\alpha_1\left[F+\left(P_\text{heat}-R\right)F^{\prime}\right]_{R=R_*}$, and $F^\prime=\mathrm{d}F/\mathrm{d}P_t$. The PWSO density limit $n_c$ can be obtained as follows
\begin{equation}
    \label{eq:threshold}
    n_e\le n_c=\frac{2D_\perp}{f_\text{ion}\lambda R_\text{imp,coe}a}\frac{1}{F+ {P_tF^{\prime}}}
\end{equation}
The power dependence of the density limit above mainly comes from the function $F(P_t)$, where $P_t$ relates to the input power $P_\text{heat}$ through the power balance equation in Eq.~(\ref{eq:pbe0}). 
 
For simplicity, a generic power function is assumed first, i.e. $F\left(P_t\right)=\alpha_2 P_t[MW]^{\mu-1}$, which yields a specific form of power dependent density limit as
\begin{equation}
    \label{eq:nc_P}
    \begin{aligned}
        n_e & \le n_c=\frac{2D_\perp}{f_\mathrm{ion}\lambda R_\text{imp,coe}a}\frac{1}{\alpha_2\mu\left(P_\text{heat}-R_*\right)^{\mu-1}}=\frac{2D_\perp}{f_\mathrm{ion}\lambda R_\text{imp,coe}a}\frac{P_\text{heat}^{1-\mu}}{\alpha_2\mu\left(\frac{\mu}{\mu+1}\right)^{\mu-1}}
    \end{aligned}
\end{equation}
where the steady-state relations
\begin{equation}
    R_*=\alpha_1\alpha_2P_t^\mu,\qquad
    R_*=\frac{P_\text{heat}}{\mu+1},\qquad P-R_*=\frac{\mu}{\mu+1}P_\text{heat}
    \label{eq:steady}
\end{equation}
are used. Considering the sputtering of tungsten by deuterium, $R_\text{imp,coe}=10^{-30}\mathrm{Wm}^{3}$ is estimated based on the simulation results using the FLYCHK code \cite{FLYCHK2008}, the coefficient $\alpha_2$ is assumed to be $10^{-5}$, $f_\text{ion}=5\times10^{-2}$, $\lambda=10^{-2}\mathrm{m}$, $a=0.5\mathrm{m}$, and $D_\perp=1\mathrm{m^2s^{-s}}$.
According Eq.~(\ref{eq:nc_P}), the heating power dependence of density limit $n_c$, which depends on the value of exponent $\mu$, are shown in Fig.~\ref{fig:nc_P_simple}. The density limit increases (decreases) with the heating power as $\mu<1$ ($\mu>1$), and becomes less sensitive to heating power at higher power level.
In reality, $\mu$ depends on the sputtering property and the relation between the target region plasma temperature $T_t$ and the power deposition on target $P_t$. Several different $\mu$ values observed in experiments can be thus determined based on the specific machine regimes, which we evaluate further in next section.

\section{Power dependence of density limit for specific targets}
\label{sec3}
In this section, more realistic functions $F(P_t)$, which is determined by the specific sputtering property in terms of the yield function, and the relation between $P_t$ and $T_t$, are evaluated to obtained more realistic dependence of density limit on edge power $P_t$ and total heating power $P_\text{heat}$.

The relation between edge temperature $T_t$ and edge power $P_t$ can be established theoretically based on 0D particle and energy balance equation (Eq.~4.38 of reference \cite{stangeby2000})
\begin{equation}
    \label{eq:stangebysimplePtTt}
    kT_\mathrm{t}=P_\mathrm{t}\frac{\tau_p}{{n}_eV}\frac{A_{\Gamma\|}}{\gamma A_{q\|}}
\end{equation}
where $k=1.6022 \times 10^{-19} \mathrm{J/eV}$ is the Boltzmann constant, $V=2\pi^2 \kappa R a^2$ is the volume, $R$ and $a$ are the major and minor radii respectively, $\kappa$ is the elongation, $\gamma$ the sheath heat transmission coefficient, $A_{\Gamma\|}$ and $A_{q\|}$ are the wet areas of particle and energy flux respectively, and $\tau_p$ is the particle confinement time for electrons. The above relation can be integrated with the sputtering function (Eq.~(\ref{eq:ITt})) to derive the specific function $F(P_t)$ (Eq.~(\ref{eq:FPtTt})) and the associated power dependence of density limit. 

We first take the specific $\tau_p$ scaling based on the JET tokamak experimental data 
$\tau_p[s]\approx1.3\times10^{14}R[m]a[m]^2\left({n}_e[m^{-3}]\right)^{-0.8}$ \cite{stangeby2000,tsuji1991,Gohen1987}.
Assuming $\kappa=1.5$, $\gamma=7$, $A_{\Gamma\|}/A_{q\|}=1$, Eq.~(\ref{eq:stangebysimplePtTt}) can be rewritten as
\begin{equation}
    \label{eq:PtTt_JET}
    T_{t}[\mathrm{eV}]=3.9\times 10^{30}P_{t}[\mathrm{W}]\left({n}_{e}[\mathrm{m}^{-3}]\right)^{-1.8}
\end{equation}
Taking the sputtering of tungsten by deuterium as an example, the function $F(P_t)$ is evaluated based on Eq.~(\ref{eq:ITt}), Eq.~(\ref{eq:FPtTt}) and Eq.~(\ref{eq:PtTt_JET}).
Then the dependence of the density limit on the heating power is obtained through Eq.~(\ref{eq:threshold}) and Eq.~(\ref{eq:pbe0}) as shown in Fig.~\ref{fig:nc_Rimp_Pheat}. There are two branches, namely, the high-$T_t$ (>35.5~eV) and the low-$T_t$ (<18~eV) branches, which are fitted to the scalings $n_c\propto P_\text{heat}^{0.255}$ and $n_c\propto P_\text{heat}^{0.744}$ respectively.
The high-$T_t$ branch scaling $n_c\propto P_\text{heat}^{0.255}$ is consistent with the experimental density limit scaling of the tungsten-wall device ASDEX-U $n_c\propto P_\text{heat}^\alpha q^{\beta}_{cyl}B^{\gamma}_{tor}$, where $\alpha \in (0.1,0.5)$~\cite{Manz2023}.
The low-$T_t$ density limit scaling, which is in the density free regime in PWSO theory, is relatively higher than the density limit in above and other experiments, such as JET experiment data \cite{tanga1987}. The ratio of impurity radiation to heating power $R_\text{imp}/P_\text{heat}$ for the corresponding two density limit branches also lie on two branches (Fig.~\ref{fig:nc_Rimp_Pheat}), where the lower branch of $R_\text{imp}/P_\text{heat}$ indicates that a relatively small amount of impurity radiation can also lead to density limit disruption. Moreover, in between these two regimes, an intermediate interval exists where $n_c$ decreases with $P_\text{heat}$.

Similarly, considering the parameters of ASDEX and W7-AS devices, $T_{t}[\mathrm{eV}]=2.5\times 10^{55}P_{t}[\mathrm{W}]\bar{n}_{e,\mathrm{main}}^{-3}[\mathrm{m}^{-3}]$ and $T_{t}[\mathrm{eV}]=8.0\times 10^{34}P_{t}[\mathrm{W}]\bar{n}_{e,\mathrm{main}}^{-2}[\mathrm{m}^{-3}]$ can be obtained by assuming modified $\tau_p$ scaling based on JET's (Eq.~\ref{eq:PtTt_JET}) $\tau_p[s]\approx 8.3\times 10^{38}R[m]a[m]^2n[m^{-3}]^{-2}$ and $\tau_p[s]\approx 2.66\times 10^{18}R[m]a[m]^2n[m^{-1}]^{-1}$ respectively for these two devices.
Considering the sputtering of deuterium on boron~\cite{Staebler1993ComparisonOD} and $D_\perp=10^{-2}\mathrm{m^2/s}$, the theoretical predictions are quantitatively consistent with the ASDEX and W7-AS experimental data from Fig.~20 of \cite{Greenwald_2002}, the same as Fig.~1 of \cite{Staebler1993ComparisonOD},  and the experimental data are located in high-$T_t$ and low-$T_t$ regime of PWSO predictions respectively, as shown in Fig.~\ref{fig:asdex-w7as}.

The particle confinement time scaling may differ in other and future tokamak devices \cite{tsuji1991,Lomanowski2022}. For instance, the particle confinement time $\tau_p$ based on ohm discharges of DIVE tokamak scales like $\tau_p\propto \sqrt{q_a}\bar{n}_e$, where $q_a$ is the safety factor at edge and $\bar{n}_e[10^{20}m^{-3}]$ is the line averaged electron density \cite{tsuji1991,divegroup1978}. Thus the corresponding power dependence of density limit may vary as well.

\section{Effects of impurity radiation from non-sputtered impurities}
\label{sec4}
In the above analysis, only the impurity radiation from the sputtered impurities is considered. In this section, the radiation due to non-sputtered impurities is also taken into account. The relation between the PWSO density limit $n_c$ and the power $P_t$ in previous section remains the same as in Eq.~(\ref{eq:threshold}). Compared to the case in absence of radiation effects from non-sputtered impurities, only the power balance becomes different. 
Additional heating power $\Delta P_{heat}$ is required to compensate the radiation power from non-sputtered impurities for the same density limit $n_c$ and the power $P_t$ since
\begin{align}
    P_\text{heat, without non-sputtered}&=P_t+R_\text{sputtered}\\
    P_\text{heat, with non-sputtered}&=P_t+R_\text{sputtered}+R_\text{non-sputtered}\\
    \Delta P_{heat}=R_\text{non-sputtered}&=n_c^2f_\text{non-sputtered}R_\text{coe,non-sputtered}
\end{align}
where $f_\text{non-sputtered}=n_\text{non-sputtered}/n_e$ is the fraction level of non-sputtered impurities. Take the non-sputtered carbon fraction $f_\text{non-sputtered}=5\%$ as an example, the heating power dependencies with and without consideration of radiation due to non-sputtered impurity are shown in Fig.~\ref{fig:nc_Pheat_with-out-non-sput_121624}. Non-sputtered impurity radiation lowers the low-$T_t$ branch of the density limit, whereas the high-$T_t$ branch remains nearly the same.

\section{Effects of fusion reactions}
\label{sec5}
In the above discussion, all heating power $P_\text{heat}$ comes from
outside the plasma. For current and future reactors with
burning plasma, the effects of fusion reactions, including
the $\alpha$ particle heating and radiation due to helium ash
may play an important role. In this section, we consider and evaluate the
effect of fusion reactions on the heating power dependence of density limit. Similar to the introduction of non-sputtered impurities, the relation between the PWSO density limit $n_c$ and the power $P_t$ remain unchanged for any given specific $F(P_t)$ as in Eq.~(\ref{eq:threshold}), and only the power balance equation requires modifications due to contributions from fusion reaction products. 

The first thing is to assess the fraction of $\alpha$ particles $f_\alpha=f_\text{He}=n_\text{He}/n_e$. Here we assume an instantaneously and complete thermalization of $\alpha$-particles out of fusion reactions. The fraction level of sputtered impurities is denoted by $f_\mathrm{imp}={n_\mathrm{imp}}/{n_e}$.
The conservation of Helium particle can be represented as
\begin{equation}
    \dfrac{n_\text{He}}{\tau_\text{He}}=S_\text{He}, \quad  S_\text{He}=\langle\sigma v\rangle n_Tn_D
    \label{eq:ConsHe}
\end{equation}
where $n_D$ and $n_T$ are the density of deuterium and tritium, $\langle\sigma v\rangle$ is the fusion reaction rate taken from \cite{Mavrin2018}. Combined with the quasi-neutral condition
\begin{equation}
    n_D+n_T+n_\mathrm{He}Z_\mathrm{He}+n_\mathrm{imp}Z_\text{imp}=n_e
\end{equation}
where $Z_\text{He}$ is the charge number of helium ions, $Z_\text{imp}$ is the charge number of sputtered impurity ions. The fraction of Helium $f_\text{He}$ in a plasma with equal proportions of D and T can be determined using Eq.~(\ref{eq:ConsHe}) as following
\begin{equation}
    \label{eq:nhe}
    n_ef_\mathrm{He}=\tau_\text{He}\frac{\left<\sigma v\right>n_e^2}{4}(1-f_\mathrm{He}Z_\mathrm{He}-f_\mathrm{imp}Z_\mathrm{imp})^2
\end{equation}
It can be shown there is and only one solution for $f_\text{He}$ in interval (0,1), which is
\begin{equation}
    f_\mathrm{He}=\frac{- B - \sqrt{B^2-4AC}}{2A}
    \label{eq:fhesol}
\end{equation}
where 
$$
A=\frac{Z_\mathrm{He}^2\tau_\text{He}\left<\sigma v\right>n_e}{4}, B=-\frac{\tau_\text{He}\left<\sigma v\right>n_eZ_\mathrm{He}\left(1-f_\mathrm{imp}Z_\mathrm{imp}\right)}{2}-1,
C=\frac{{\tau_\text{He}\left<\sigma v\right>n_e}\left(1-f_\mathrm{imp}Z_\mathrm{imp}\right)^2}{4}
$$
The dependence of helium fraction $f_\text{He}$ on the plasma temperature $T$ and impurity fraction $f_\text{imp}$ determined from Eq.~(\ref{eq:fhesol}) can be evaluated, given the confinement time $\tau_\text{He}$ one second, which is the order required for satisfying the Lawson criterion, in the range of number density and temperature considered in Fig.~\ref{fig:fHe_fimp_T_102224}.

The second step is to evaluate the power of $\alpha$ particle heating $P_\alpha$ and helium radiation $R_\text{He}$ at the PWSO density limit. The helium radiation power is 
\begin{equation}
    \label{eq:alpha}
    R_\text{He}=f_\text{He}n_c^2R_\text{coe,He}, \quad P_\alpha=\frac{\langle \sigma v\rangle}{4} U_\alpha n_c^2(1-f_\mathrm{He} Z_\text{He}-f_\mathrm{imp} Z_\text{imp})^2  
\end{equation}
where $n_c$ is the density limit as in Eq.~(\ref{eq:threshold}), $f_\text{He}$ can be obtained through Eq.~(\ref{eq:fhesol}) for the given plasma temperature $T$ and the fraction level of sputtered impurity $f_\text{imp}$ governed by Eq.~(\ref{eq:fimpPheat}). 
Considering the JET specific scaling for $\tau_p$ in Eq.~(\ref{eq:PtTt_JET}) and the corresponding function $F(P_t)$ in Section \ref{sec3}, and calculating the external heating power through the power balance equation as follow.
\begin{equation}
    \label{eq:0Dbalance}
    P_\text{heat}+P_\alpha-R_\text{imp,sputtered}-R_\text{He}=P_t
\end{equation}
The density limit power dependence with and without effects of fusion reactions shown in Fig.~\ref{fig:nc_Pheat_with-out-fusion_121624} demonstrate that fusion reactions will enhance the density limit at the same external heating power, which is a result of $\alpha$ particle heating. The effects of fusion reactions are able to enhance the density limit and reduce the needed external heating power to reach the same density limit on the low-$T_t$ branch. However, these effects are very weak on the high-$T_t$ branch. A related discussion about requirements of achieving burning plasma condition as predicted using the 0D PWSO model is presented in Section \ref{sec:A1}.

\section{Summary}
In this work, the heating power dependence of density limit in tokamak is explored by considering the plasma-wall interaction through the plasma-wall self-organization theory. The results shows that there are two density limit regimes, low-$T_t$ and high-$T_t$ regions, corresponding to the density-free and density limit regime of the PWSO theory \cite{escande2022,jxliu2023}, respectively. In particular, in the high-$T_t$ region, the density limit power dependence is about $n_c\propto P_\text{heat}^{0.255}$ for the sputtering of tungsten from deuterium and the particle confinement time scaling $\tau_p[s]\approx1.3\times10^{14}R[m]a[m]^2\left(\bar{n}_\mathrm{main}[m^{-3}]\right)^{-0.8}$ based on the experimental results on JET tokamak. This density limit scaling agree well with the other experimental and theoretical power dependent density limit scaling, especially those listed in Table.~1 of \cite{Manz2023}. The low-$T_t$ region's density limit scaling is about $n_c\propto P_\text{heat}^{0.744}$ which is much higher than the above DL scaling. Besides, the predictions of this model quantitatively matches experimental trends across devices, including ASDEX-U and W7-AS.
The effects of radiation from non-sputtered impurities, and fusion products on the power dependence of density limit are also evaluated. The burning plasma condition under the density limit status is explored and predicted considering the plasma wall interaction. These results underscore the necessity of regulating plasma-wall interactions and confinement properties to achieve higher density limit, offering a pathway toward optimizing performance in next-step fusion reactors.

In future work, we plan to extend the zero-dimensional analysis presented here to higher-dimensional studies using the PWSO theory in 1.5-dimensional integrated simulations, where the effects of various heating schemes, transport mechanisms, and the influence of 1D spatial profiles for electron density, plasma current, and temperature can be taken into account.
In addition, cross-machine experiments are proposed on EAST and HL-3 to validate the power dependence scaling of density limit and the role of plasma-wall interaction.

\section*{Acknowledgment}
This work is supported by the National MCF Energy R\&D Program of China under Grant No.~2019YFE03050004 and the U.S. Department of Energy Grant No.~DE-FG02-86ER53218. The computing work in this paper is supported by the Public Service Platform of High Performance Computing by Network and Computing Center of HUST.

\appendix
\section{Burning plasma condition in 0D PWSO model}
\label{sec:A1}
The energy gain factor $Q=5P_\alpha/P_\text{heat}$ is evaluated for the case in Section \ref{sec5}, and it turns out that the $Q$ value is always much lower than 1 in the temperature ranges in Fig.~\ref{fig:Q_VS_010725}, which means the energy break-even ($Q=1$) can not be achieved for the JET $\tau_p$ scaling in Eq.~(\ref{eq:PtTt_JET}). For a given plasma temperature and density, the $\alpha$ heating power $P_{\alpha}$ is fixed as in Eq.~\ref{eq:alpha}. To reach a higher energy gain factor $Q$, the relation between $n_e$, $T_t$ and $P_t$ (Eq.~\ref{eq:PtTt_JET})) needs modifications,such as increase the coefficient in front of $P_t$ or increase the exponent of $n_e$, which would allow a lower power $P_t$ for given electron density $n_e$ and temperature $T_t$. This is essentially a requirement for the scaling of particle confinement $\tau_p$. 
Taking the following relation  as an example
\begin{equation}
    \label{eq:modifiedtaup}
    T_{t}[\mathrm{eV}]=3.9\times 10^{30}P_{t}[\mathrm{W}]\left({n}_{e}[\mathrm{m}^{-3}]\right)^{-1.745}
\end{equation}
which introduce a slightly different $n_e$ scaling from that in Eq.~\ref{eq:PtTt_JET}.
This scaling will lead to a much higher energy gain factor as shown in Fig.~\ref{fig:Q_VS_010725}. 

\makeatletter
\renewcommand{\thesection}{\@arabic\c@section}
\makeatother
\renewcommand{\thefigure}{\arabic{figure}}

\section*{References}
\bibliographystyle{unsrt}
\bibliography{references}

\newpage
\begin{figure}[htbp]
    \centering
    \includegraphics[width=\linewidth]{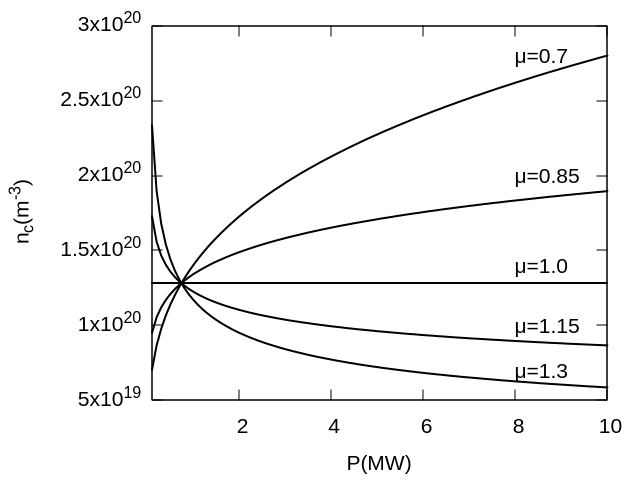}
    \caption{Density limit $n_c$ as functions of the heating power $P$ with various $\mu$ values}.
    \label{fig:nc_P_simple}
\end{figure}

\newpage
\begin{figure}[htbp]
    \centering
    \includegraphics[width=\linewidth]{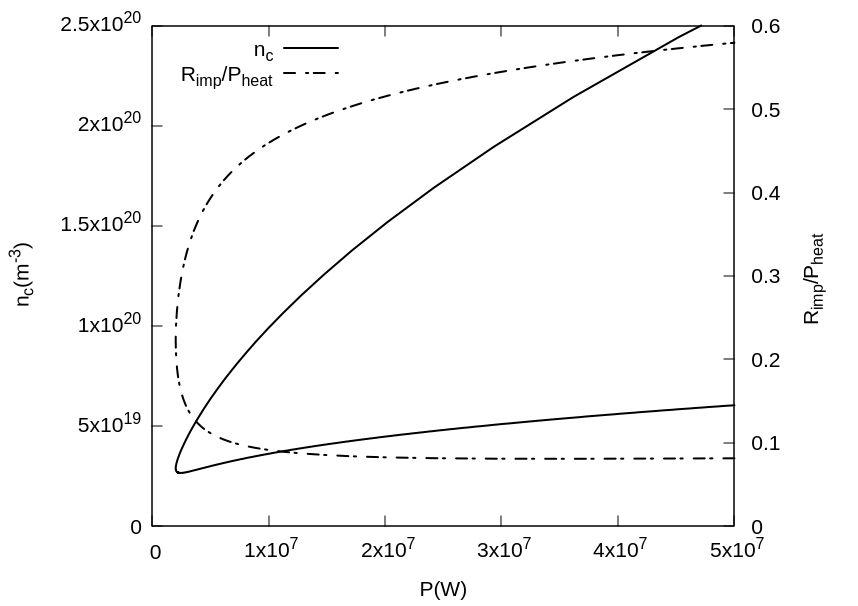}
    \caption{Density limit $n_c$ (solid line) and the radiation to heating power ratio ${R_\mathrm{imp}}/{P_\mathrm{heat}}$ (dashed line) as functions of the heating power $P_\mathrm{heat}$
    with the power and electron density dependence of $\tau_p$ in JET experiments as shown in Eq.~(\ref{eq:PtTt_JET}).}
    \label{fig:nc_Rimp_Pheat}
\end{figure}

\newpage
\begin{figure}
    \centering
    \includegraphics[width=\linewidth]{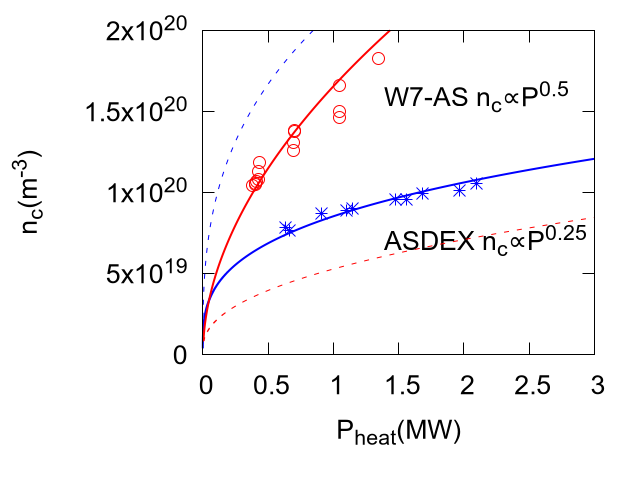}
    \caption{Density limit $n_c$ as functions of heating power $P_\text{heat}$ from PWSO theory and the experimental data of ASDEX tokamak and W7-AS stellarator considering the sputtering of deuterium on boron and $D_\perp=10^{-2}\mathrm{m^2/s}$. The experimental and devices' data is taken from \cite{Greenwald_2002,Staebler1993ComparisonOD}.}
    \label{fig:asdex-w7as}
\end{figure}

\newpage
\begin{figure}
    \centering
    \includegraphics[width=\linewidth]{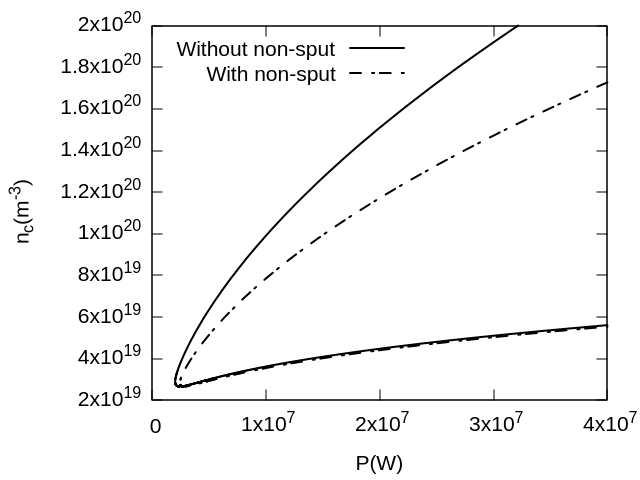}
    \caption{Density limit $n_c$ (solid line) as functions of heating power $P_\mathrm{heat}$ with (dashed line) and without (solid line) considering the radiation effect due to non-sputtered impurity.}
    \label{fig:nc_Pheat_with-out-non-sput_121624}
\end{figure}

\newpage
\begin{figure}[htbp]
    \centering
    \includegraphics[width=0.8\linewidth]{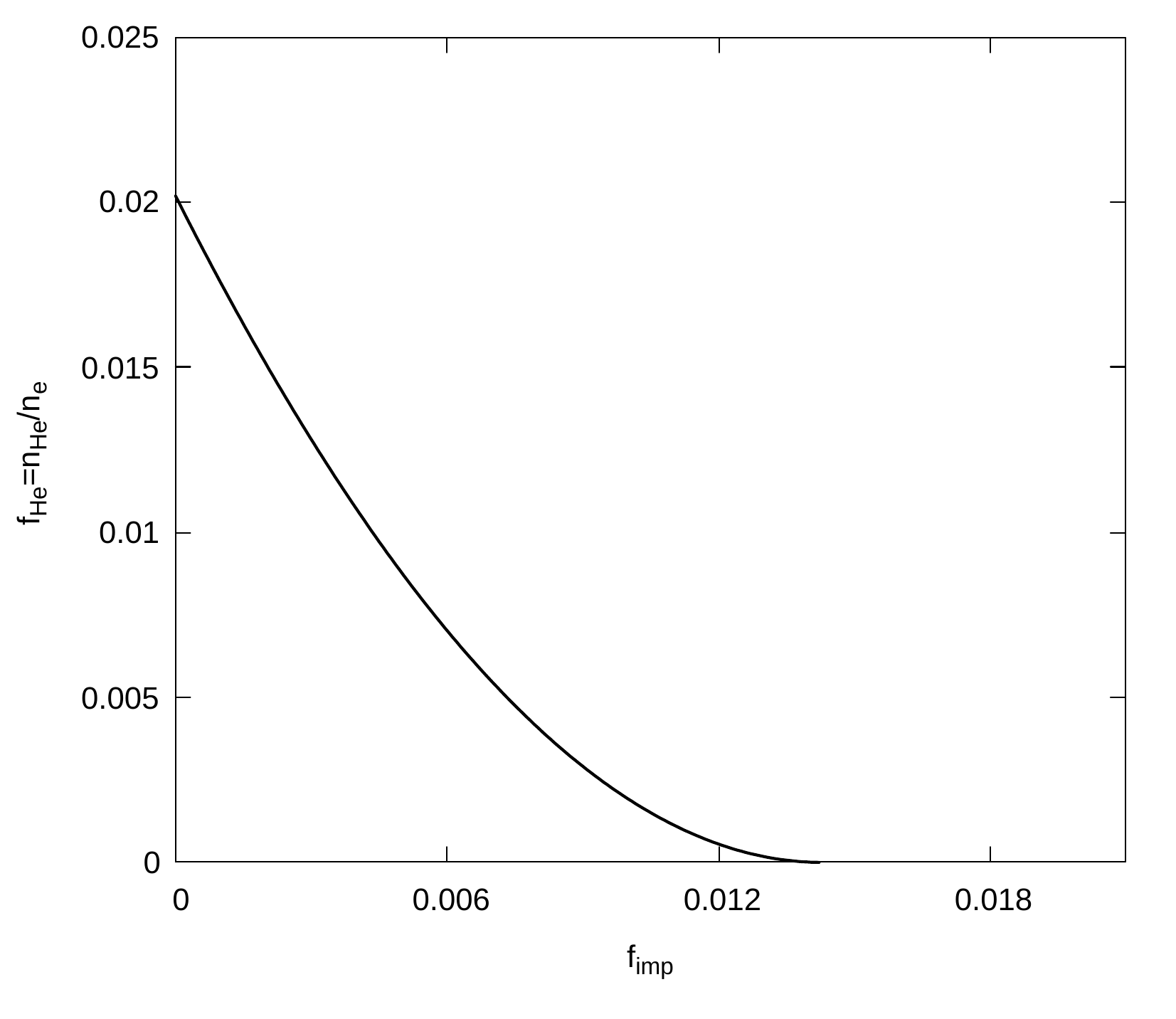}
    \includegraphics[width=0.8\linewidth]{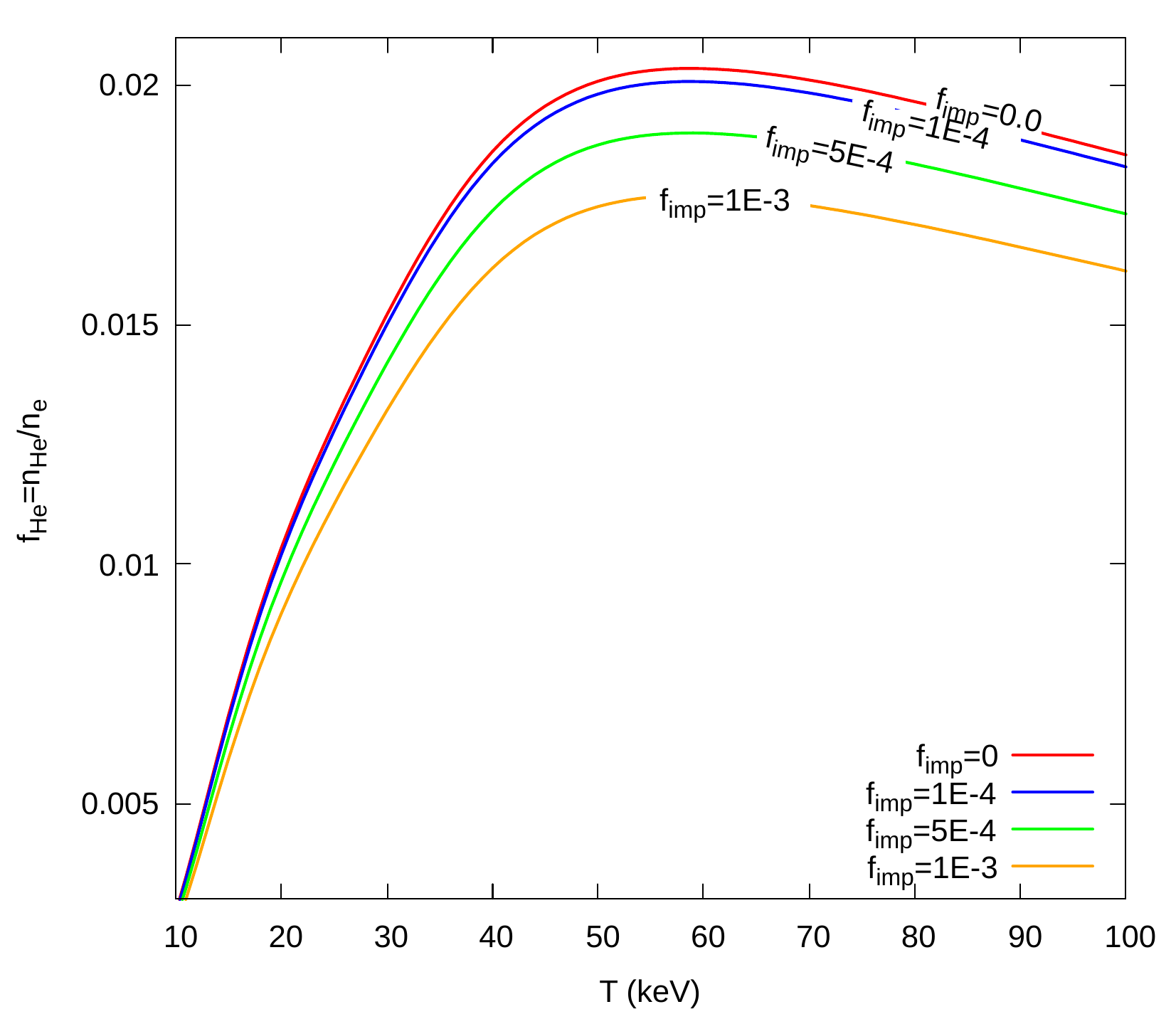}
    \caption{The helium fraction $f_\text{He}$ as a function of (top) impurity level $f_\text{imp}$ at the core plasma temperature $T=51.8$ keV and as functions of (bottom) the core plasma temperature $T$ with various impurity levels at electron density $n_e=10^{20}\text{m}^{-3}$ and impurity's charge number $Z_\text{imp}=70$.}
    \label{fig:fHe_fimp_T_102224}
\end{figure}

\newpage
\begin{figure}[htbp]
    \centering
    \includegraphics[width=\linewidth]{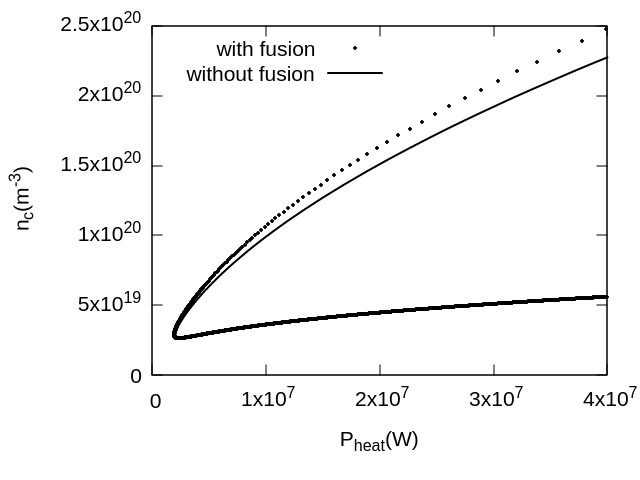}
    \caption{Density limit as functions of heating power $P_\text{heat}$ (points) with and (solid line) without effects of fusion reactions at plasma temperature $T=50$keV.}
    \label{fig:nc_Pheat_with-out-fusion_121624}
\end{figure}

\newpage
\begin{figure}[htbp]
    \centering
    \includegraphics[width=0.9\linewidth]{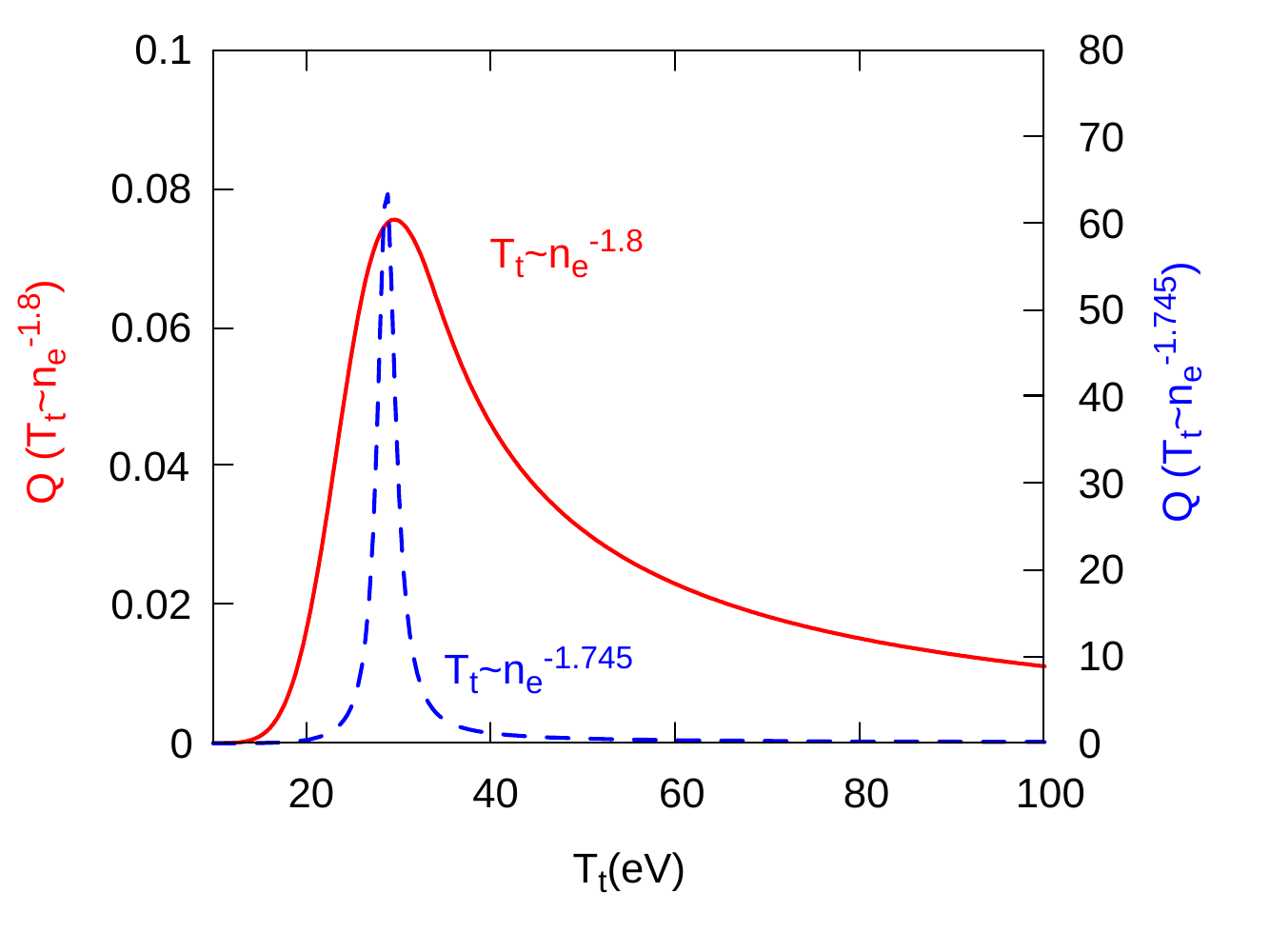}
    \includegraphics[width=0.9\linewidth]{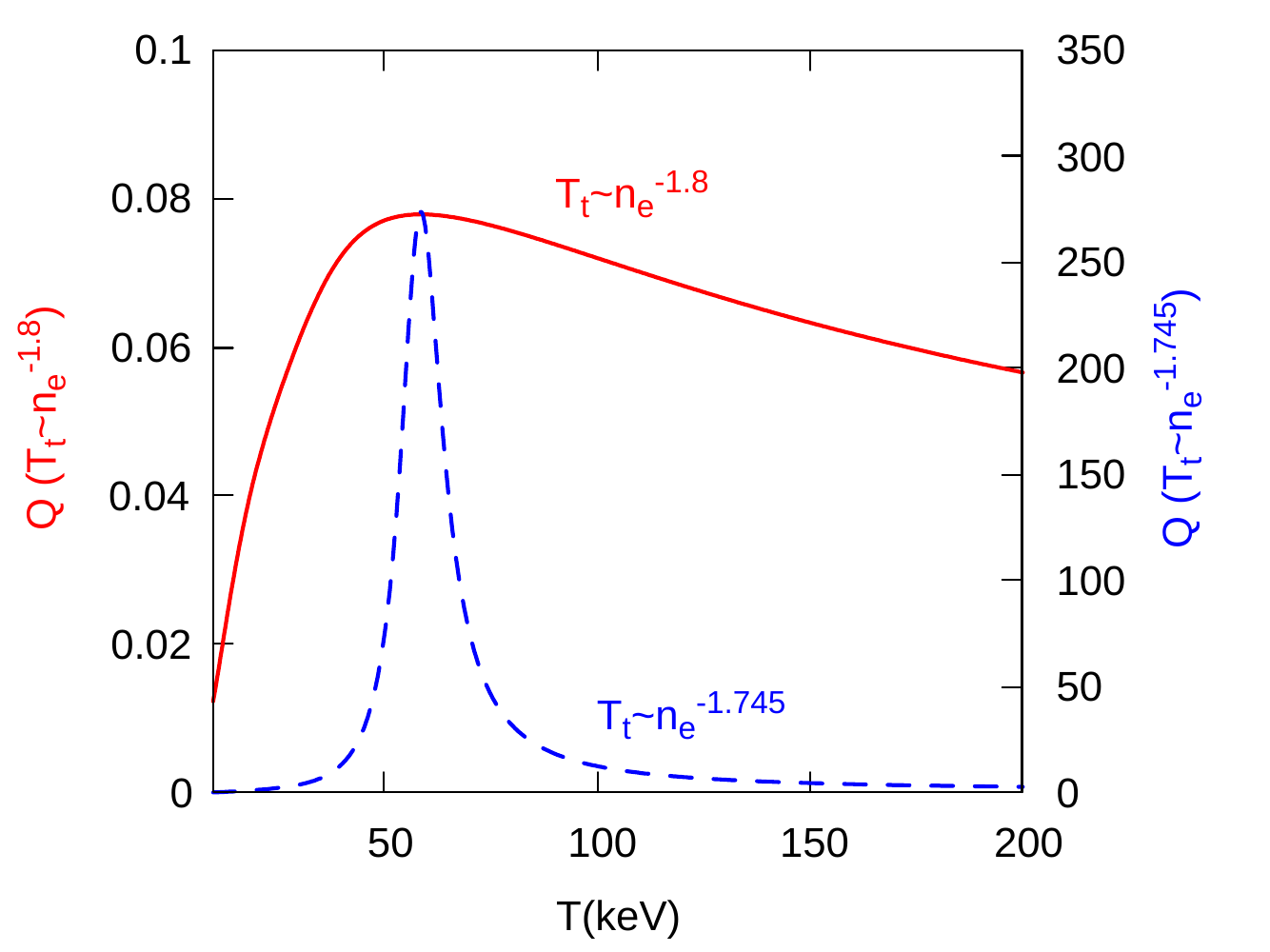}
    \caption{Energy gain factor $Q$ as a function of (top) the target region plasma temperature $T_t$(eV) at the core plasma temperature $T=45$ keV and as a function of (bottom) the core plasma temperature $T$(keV) at $T_t=30$ eV for (red solid line) the JET $\tau_p$ scaling in Eq.~(\ref{eq:PtTt_JET}) and (blue dashed line) the modified $\tau_p$ scaling in Eq.~(\ref{eq:modifiedtaup}).}
    \label{fig:Q_VS_010725}
\end{figure}

\end{document}